\documentstyle[12pt,titlepage]{article}
\newcommand{\Eqref}[1]{Eq.(\ref{#1})}
\newcommand{\Eqsref}[2]{Eqs.(\ref{#1},\,\ref{#2})}

\newcommand{\Figref}[1]{Fig.(\ref{#1})}

\newcommand{\bea}{\begin{eqnarray}}
\newcommand{\eea}{\end{eqnarray}}
\newcommand{\be}{\begin{equation}}
\newcommand{\ee}{\end{equation}}
\newcommand{\bc}{\begin{center}}
\newcommand{\ec}{\end{center}}
\newcommand{\ba}{\begin{array}}
\newcommand{\ea}{\end{array}}
\newcommand{\bfig}{\begin{figure}}
\newcommand{\efig}{\end{figure}}
\newcommand{\non}{\nonumber}

\newcommand{\inv}[1]{\frac{1}{#1}}
\newcommand{\del}{\partial}

\newcommand{\ie}{\mbox{\sl i.e. }}

\newcommand{\np}[3]{{\it  Nucl. Phys. }{{\bf #1} {(#2)} {#3}}}
\newcommand{\pr}[3]{{\it Phys. Rev.}{{ \bf #1} {(#2)} {#3}}}

\newcommand{\prd}[3]{{\it  Phys. Rev. D} {{\bf #1} {(#2)} {#3}}}

\newcommand{\pl}[3]{{\it  Phys. Lett. }{{\bf #1} {(#2)} {#3}}}

\newcommand{\sovnp}[3]{{\it Sov. J. Nucl. Phys. }{{\bf #1} {(#2)} {#3}}}

\newcommand{\th}{\theta}
\newcommand{\trmr}{\frac{T_R}{m_R}}
\newcommand{\tmr}{\frac{T}{m_R}}
\begin{document}

\input FEYNMAN
\large
\thispagestyle{empty}
\begin{flushright} NORDITA-92/63  P \\

                   hep-ph/9209282  \\
                   Sept 1992  \end{flushright}
\bc
{\LARGE\bf Temperature Renormalization Group \vspace{4mm}\\ and Resummation}
\ec
\vspace*{1cm}
\bc  {\Large Per Elmfors}\

\footnote{ Permanent address: NORDITA, Blegdamsvej 17, DK-2100 Copenhagen,
	  Denmark,\\ E-mail: elmfors@nordita.dk} \vspace*{1cm} \\
	The Theoretical Physics Institute \\
	University of  Alberta\\
	Edmonton \\
	Alberta T6G 2J1\\

	Canada\ec
\vspace*{2cm}
\bc
{\bf Abstract} \\
\ec {\normalsize
The temperature renormalization group equation (TRGE) is compared with a
diagrammatic
expansion for the $(\phi^4)_4$-theory. It is found that the one-loop TRGE
resums the
leading powers of temperature for the effective mass. A two-loop contribution
to TRGE is
required to do the leading resummation for the coupling constant. It is also
shown that the
higher order TRGE resums subleading powers of temperature.}
\newpage
%
% -----------------------------------------------------------------
\normalsize
\setcounter{page}{1}
\section{Introduction}
The temperature renormalization group equation \cite{matsumotonu84} (TRGE) has
been
used by several authors to study the high temperature behaviour of different
theories
\cite{fujimotoiny86,funakubos87,fujimotoy88,nakkagawany88,elmfors92}. With few
exceptions \cite{funakubos87} it has not been clear exactly what kind of
improvement the
TRGE implies. In some cases \cite{fujimotoiny86,elmfors92}  a 1-loop
approximation of the
effective potential was used to determine the renormalization flow of the mass
and
coupling constant at zero momentum. This gives the first correction in $\hbar$
but not
necessarily the leading correction in $T/m$ as we shall see. The purpose of
this paper is to
give a more detailed discussion of the TRGE for the $(\phi^4)_4$-theory.
\\ \\
When calculating thermal corrections perturbatively in a QFT at finite
temperature it is
often found that the 1-loop result goes like a power of $T$ at high
temperature. The
correction is, therefore, always large for large enough $T$ which makes the
perturbation
approach unreliable. The self-energy, for instance, is the correction to the
effective
(mass$)^2$ of a  particle surrounded by a heat-bath. It gets a correction going
like $T^2$ at
high $T$ in the $(\phi^4)_4$-theory. If the quasi-particles in the thermal
environment have
a mass  very different from the zero temperature one, it is clear that it is
not a good
approximation to expand around the zero temperature excitations. On the other
hand,
expanding around quasi-particles at one temperature to calculate the
self-energy at a
temperature that is close has a better chance to succeed. Also, a sequence of
many small
steps in temperature is intuitively more likely to give a better result than
taking only one big
step. The situation is familiar from zero temperature calculations where the
mass can be
renormalized at different scales $\mu$. Using a mass defined at $\mu$ to
calculate the
mass at the scale $\mu'$ gives rise to $\log (\mu'/\mu)$ terms which can be
large. There
the remedy is well-known as the renormalization group which sums up small steps
in
$\mu$. The renormalization group equation (RGE) at $T=0$ can be formally
derived from
the freedom of choosing $\mu$ in the renormalization prescription. At finite
temperature,
the mass of the quasi-particles, around which to expand, can be defined at any
temperature and there is a freedom of choosing that temperature which should
not
influence on the final answer. This freedom can be used to derive the TRGE
which
implements the summation over small temperature steps \cite{matsumotonu84}.
\\ \\
The result from solving the 1-loop TRGE is certainly different from just
computing a 1-loop
correction but it is not obvious exactly what the difference is. It is known
that a
resummation of leading $\log\mu$ terms is obtained by solving the RGE at zero
temperature. Since there is a similarity between $\log\mu$ and $T$ it is
natural to guess
that the TRGE resums the leading powers of $T$. This is true for the
$(\phi^4)_4$-theory
and that is the basic result of this paper. It turns out, however, that the
TRGE can only
resum certain subdiagrams in an effective way. By effective I mean getting a
result to the
$n$-th order without doing an $n$-th order calculation. The diagrams that are
effectively
resummed are the ones containing loops that give a leading power of $T^2$ (hard
thermal
loops). These terms can also be resummed using an effective propagator with a
$T$
dependent mass \cite{dolanj74}. The relation between the two methods is
discussed in
Section \ref{effprop}.

In Section \ref{resT} the thermal $\th$ and $\beta$ functions are discussed and
their
relation to the diagrammatic expansion of $m(T)$ and $g(T)$. The TRGE to
leading order is
solved in Section \ref{i01}.
%
%% ++++++++++++++++++++++++++++++++++++++++++++++++++++
\section{Resummation of leading powers of $T$}
\label{resT}

%
%''''''''''''''''''''''''''''''''''''''''''''''''''''''
\subsection{General ideas}
\label{gen}

Our goal is to calculate the mass and the coupling as functions of the
temperature ($m(T),\
g(T)$). The meaning of $m$ and $g$ depends on the renormalization conditions
(RC). To
give a physical interpretation of $m(T)$ we define it to be the real part of
the pole of the
propagator of physical excitations. This assumes that it is meaningful to talk
about single
stable excitations which may be difficult at finite temperature when the
excitations
eventually dissipate away, but we take it as an approximation. Adding an
imaginary part to
the mass would take into account the dissipation.
\footnote{ Perturbative propagators with complex poles do not admit any
spectral
representation, which the full propagator can be shown non-perturbatively to
have (at least
at equilibrium). It may therefore be necessary to consider some generalization
of the TRGE
to the spectral function if dissipation is to be taken into account
consistently. I want to
thank Dr. Umezawa for pointing out this to me.}
 In some cases it may be needed to to consider excitations with a spectrum that
differs
drastically from the free one, but we do not discuss that here (see Section
\ref{othther}).
The coupling constant is related to experiment through some scattering process
and we
define it to be the real part of the 4-point function with external momenta
corresponding to
that scattering. For simplicity we assume that the momentum depends only on the
mass
and some parameters that are kept fix, such as scattering angles. The
renormalization
point is then denoted by $p_{iR}=p_i(m_R)$. Thus the renormalization condition
is
\bea
\label{RC1}
	{\rm Re}\Gamma^{(2)}(p,T)|_{p^2=m^2_R,\ T=T_R}&=&0 \ ,\non\\
	{\rm Re}\Gamma^{(4)}(p_i,T)|_{p_i(m_R),\ T=T_R}&=&-g_R(T_R) \ .
\eea
In addition, the wave function has to be renormalized and we choose
\be
\label{RC2}
	\frac{\del}{\del p_0^2}{\rm Re}
	\Gamma ^{(2)}|_{p^2=m^2_R,\ T=T_R}=-1\ ,
\ee
as RC.
Other conditions (like the MS scheme) can be imposed if the only purpose is to
eliminate
the UV divergences. Here we also want to give the expansion parameters $m_R$
and
$g_R$ a physical meaning.
\\ \\
{}From the RC the corresponding $\th$, $\beta$ and $\gamma$ functions can be
defined
\cite{matsumotonu84}
\footnote{ The definition of $\th$, $\beta$ and $\gamma$ differs from
Ref.\cite{matsumotonu84} by factors of $T_R$ and $m_R$.}.
\bea
\label{bfcts}
	\th(g_R,m_R,T_R)&=&-\inv{2m_R}\frac{\del}{\del T}
	{\rm Re}\Gamma^{(2)}(p_R,T)|_{T_R} \ ,\non\\
	\beta(g_R,m_R,T_R)&=&4\gamma g_R-\frac{\del}{\del T}
	{\rm Re}\Gamma^{(4)}(p_{iR},T)|_{T_R} \ ,\non\\
	\gamma(g_R,m_R,T_R)&=&-\inv{2}\frac{\del}
	{\del T}\frac{\del}{\del p_0^2}
	{\rm Re}\Gamma^{(2)}(p,T)|_{p_R,\ T_R} \ .
\eea
The mass and coupling, defined through \Eqref{RC1}, are obtained by integrating
$\th$
and $\beta$,
\be
\label{TRGE}
	\left\{\ba{lcr} \frac{dm}{dT}&=&\th(g,m,T) \non\\
		\frac{dg}{dT}&=&\beta(g,m,T) \\
	\ea\right.
	=>
	\left\{\ba{lcr} m&=&m(T;g_R,m_R,T_R) \non\\
	g&=&g(T;g_R,m_R,T_R) \\
	\ea\right. \ ,
\ee
and we get in this way the zero of Re$\Gamma^{(2)}(p,T)$ and the value of
Re$\Gamma^{(4)}(p_R,T)$ as functions of $g_R$, $m_R$, $T_R$ and $T$.

The same functions could in principle be calculated by the diagrammatic
perturbation
series as an expansion in $g_R$. The idea here is to compute $\th$ and $\beta$
perturbatively and then solve the non-linear ordinary differential equations
\Eqref{TRGE}
which are non-perturbative relations. To study the relation between the two
approaches it
is convenient to use the invariance under a change of the initial conditions to
write the
linear partial differential equations
\bea
\label{PDE}
	\left( \frac{\del}{\del T_R} +\th(g_R,m_R,T_R)\frac{\del}{\del m_R}
	+\beta(g_R,m_R,T_R)\frac{\del}{\del g_R}\right)
	m(T;g_R,m_R,T_R)&=&0 \ ,\non\\
	\left( \frac{\del}{\del T_R} +\th(g_R,m_R,T_R)\frac{\del}{\del m_R}
	+\beta(g_R,m_R,T_R)\frac{\del}{\del g_R}\right)
	g(T;g_R,m_R,T_R)&=&0 \ .
\eea
\\ \\
If now $\th$ and $\beta$ are computed to finite order in $g_R$ what do the
corresponding
solutions for $m$ and $g$ mean in terms of perturbation theory? To answer that
question
we make an Ansatz for the solution as a power series in $g_R$ and determine the
coefficients. Before doing that we shall take a closer look at the leading $T$
behaviour of
$\th$ and $\beta$. We must also be aware that, by making the Ansatz above, we
assume a
power series expansion of $m$ and $g$ around $g_R=0$.
%
% '''''''''''''''''''''''''''''''''''''''''''''''''''''''''''
\subsection{Power series for $\th$ and $\beta$}
\label{powtb}

We are interested in the leading $T$ dependence of $\th$ and $\beta$ at each
order in
$g_R$, so we have to find the diagrams that contribute most. According to
\Eqref{bfcts} the
$\th$ and $\beta$ functions are obtained by taking suitable $T$ and $p$
derivatives of
Re$\Gamma^{(2)}$ and Re$\Gamma^{(4)}$ and putting \mbox{$p=p_R,\ T=T_R$}
afterwards. The Re$\Gamma^{(2)}$ and Re$\Gamma^{(4)}$ should be renormalized
with
the physical self-consistent RC in \Eqsref{RC1}{RC2}. This turns to be crucial.

The dominant diagram at high $T$ is the tadpole correction to the mass which
gives a
factor $T^2$. All other loops give a factor $T$. Therefore, the dominant
diagrams to each
order in $g_R$ are the ones with as many tadpole subdiagrams as possible. This
is true for
$\Gamma^{(N)}$. However, for each such tadpole there is a  corresponding
diagram with a
counterterm defined at $T_R$ (see \Figref{tadpole}).
%

% ** Tadpole **
\bfig
\begin{picture}(20000,8000)(-10000,0)
	\thicklines
	\drawline\fermion[\E\REG](1000,3500)[7000]
	\global\advance \pmidy by 2000
	\put(\pmidx,\pmidy){\circle{4000}}
	\drawline\fermion[\E\REG](15000,\pbacky)[7000]
	\put(\pmidx,\pmidy){\circle*{800}}
      \put(11000,\pbacky){\makebox(0,0){$-$}}
      \put(11000,500){\makebox(0,0){$I(T,m_R)-I(T_R,m_R)$}}
\end{picture}
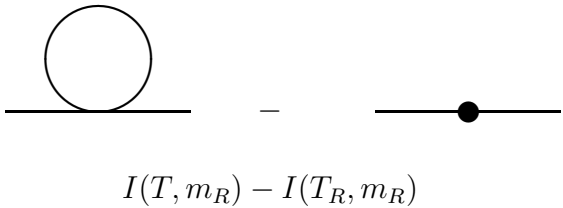
\caption{The tadpole plus its counterterm.}
\label{tadpole}
\efig
 Let us denote the tadpole contribution by $I(T,m_R)$. The sum of all diagrams
with  $n$
tadpoles and their counterterms contains a factor $(I(T,m_R)-I(T_R,m_R))^n$.
Evidently,
after taking the the $T$ derivative and putting $T=T_R$, this factor vanishes
for all
$n\geq2$. We conclude that only diagrams with at most one tadpole as subdiagram
contributes to $\th$ and $\beta$ and thus the $n$-th order term in for $\th$
and $\beta$
goes like $g_R^nT^n$ and $g_R^{n+1}T^n$ respectively.
\\ \\
We write the high $T$ expansion as
\bea
\label{bexp}
	\beta(\trmr)&=&\frac{g_R}{m_R}\sum_{k=1}^{\infty}
	g_R^k(\trmr)^k\sum_{p=0}^\infty(\trmr)^{-p}\beta_{kp} \ ,\non\\
	\th(\trmr)&=&\sum_{k=1}^{\infty}
	g_R^k(\trmr)^k\sum_{p=0}^\infty(\trmr)^{-p}\th_{kp} \ .
\eea
where $k$ is the order to which $\th$ and $\beta$ are calculated.
By looking at low order diagrams we see that $\beta_{10}$ is zero which is
important in
next section.

In addition to the terms in \Eqref{bexp} there are subleading terms like
$T^m\ln^nT$. If they
are considered to be of the same order as $T^m$ independently of $n$, they do
not affect
the ordering of terms with different powers of $T$ that is used in next
section. They also do
not occur to the leading order. Therefore, I do not write them out since they
would only
complicate the formulas.
%
%''''''''''''''''''''''''''''''''''''''''''''''''''''''''''''
\subsection{Power series for $m$ and $g$}
\label{powmg}

The purpose of this section is to relate the solutions of \Eqref{PDE} to the
perturbative
calculation of $\th$ and $\beta$. A diagrammatic expansion of $m$ and $g$
yields a power
series in $g_R$ so we make the Ansatz
\be\ba{rcl}
	m(T;g_R,m_R,T_R)&=&
	m_R\sum_{n=0}^\infty g_R^n M_n(\tmr;\trmr)\ ,\non\\
	g(T;g_R,m_R,T_R)&=&
	g_R\sum_{n=0}^\infty g_R^n G_n(\tmr;\trmr)\ ,
\ea\ee
where $M_n$ and $G_n$ only depend on $\tmr$ and $\trmr$ for dimensional
reasons. This
is the reason why we related $p_R$ to $m_R$: it reduces the number of
dimensionful
variables. Of course, we could have related $p_R$ to $T_R$ and get the same
simplification here but with an other physical interpretation of $m$ and $g$.
The leading
behaviour of $\th$ and $\beta$ in last section was considered for a $p_R$ which
does not
depend on $T_R$. We could also have treated the renormalization group equation
in both
$p_R$ and $T_R$ as proposed in \cite{matsumotonu84}.
\\ \\
Our primary interest is to see how the $T$ dependence in $\th$ and $\beta$
gives a $T$
dependence in $m$ and $g$ so we further expand $M_n$ and $G_n$
\bea
\label{MGexp}
	M_n(\tmr;\trmr)&=&\sum_{q,l}(\trmr)^q(\tmr)^lM_{nql}\ ,\non\\
	G_n(\tmr;\trmr)&=&\sum_{s,u}(\trmr)^s(\tmr)^uG_{nsu} \ ,
\eea
where the summation range for $q$, $l$, $s$ and $u$ is yet to be determined.
The
comment at the end of last section regarding possible $\ln T$ factors apply to
this
expansion as well.

{}From \Eqref{MGexp} and \Eqref{PDE} we find
\bea
\label{qMsG}
	qM_{nql}&=&-\sum_{k=1}^n\sum_{p=o}^\infty M_{n-k,q-1+p-k,l}
		[\th_{kp}(2+k-p-q-l)+\beta_{kp}(n-k)] \ , \non\\
	sG_{nsu}&=&-\sum_{k=1}^n\sum_{p=o}^\infty G_{n-k,s-1+p-k,u}
		[\th_{kp}(1+k-p-s-u)+\beta_{kp}(n-k+1)] \ ,
\eea
by identifying the coefficients of the same powers of $g_R$, $T_R$ and $T$.
The \Eqref{qMsG} determines $M_{nql}$ ($G_{nsu}$) in terms of $M_{n'ql}$
($G_{n'su}$)
where $n'<n$, and it can be applied inductively except for $q=0$ ($s=0$)
\footnote{ Again, when including $\ln T$ terms this exception is modified.}.

We are interested in the $T$ dependence but \Eqref{PDE} is a differential
equation in
$T_R$. The $T$ dependence enters through the boundary conditions which also
determines $M_{nql}$ ($G_{nsu}$) for $q=0$ ($s=0$). The boundary conditions are
\be\ba{rlc}
\label{BC}

	m(T_R;g_R,m_R,T_R)&=&m_R\ , \non\\
	g(T_R;g_R,m_R,T_R)&=&g_R\ ,

\ea\ee
or in terms of $M_n$ and $G_n$
\bea
\label{BC2}
	M_n(\trmr;\trmr)=0\ ,\ n\geq1 \ ; & & M_0=1\ , \non\\
	G_n(\trmr;\trmr)=0\ ,\ n\geq1 \ ; & & G_0=1\ .
\eea
They lead to
\be\ba{rcl}
\label{q0cond}

	M_{n0l}&=&-\sum_r' M_{n,r,l-r}\ , \non\\
	G_{n0u}&=&-\sum_v' G_{n,v,u-v}\ ,
\ea\ee
where $\sum'$ means that $r=0$ ($v=0$) is excluded.
\\  \\
We shall now determine the summation ranges for $q,l$ and $s,u$ in
\Eqref{MGexp} using
\Eqref{qMsG} and \Eqref{q0cond}. It is obvious that there is no lower limit in
general. We
define the upper limit by
\be\ba{rcl}
\label{QL}

	Q(n,l) &=& {\rm max\ } q {\rm\ for\ which\ } M_{nql} \neq 0 \non\\
 	L(n,q) &=& {\rm max\ } l {\rm\ for\ which\ } M_{nql} \neq 0 \ ,
\ea\ee
\be\ba{rcl}
\label{SU}

	S(n,u) &=& {\rm max\ } s {\rm\ for\ which\ } G_{nsu} \neq 0 \non\\
 	U(n,s) &=& {\rm max\ } u {\rm\ for\ which\ } G_{nsu} \neq 0 \ .
\ea\ee
When doing this we must be careful with the last factors in brackets in
\Eqref{qMsG} which
can be equal to zero for certain values of $k,p,n,q,l,s$ and $u$ (remember also
that
$\beta_{10}=0$). As noted before, their form is different when including
possible $\ln T$
factors. The easiest way to find $Q,L,S$ and $U$ is to do an explicit
calculation for small
$n$, guess the general form and then prove it by induction. There are some
comments
about the procedure in Appendix \ref{appA}.

The initial condition for $n=0$ is
\be\ba{rcl}
\label{n0}

	Q(0,0)=0 \ &;&
	Q(0,l\neq 0)=-\infty \ , \non\\
	L(0,0)=0 \ &;&
	L(0,q\neq 0)=-\infty \ , \non\\
	S(0,0)=0 \ &;&
	S(0,u\neq 0)=-\infty \ , \non\\
	U(0,0)=0 \ &;&
	U(0,s\neq 0)=-\infty \ . \non
\ea\ee
The result for $n\geq 1$ is
\be\ba{rcl}
\label{Q}

	Q(n,l<0)&=&2n-2 \non \\
	Q(n,0\leq l\leq 2n)&=&2(n-[\frac{l+1}{2}]) \non\\
	Q(n,l>2n)&=&-\infty \ ,
\ea\ee
\be\ba{rcl}
\label{L}

	L(n,q<0)&=&2n-2 \non \\
	L(n,0\leq q\leq 2n)&=&2(n-[\frac{q+1}{2}]) \non\\
	L(n,q>2n)&=&-\infty \ ,
\ea\ee
\be\ba{rcl}
\label{S}

	S(n,u<0)&=&2n-2 \non \\
	S(n,u=0)&=&2n-1 \non \\
	S(n,1\leq u\leq 2n-1)&=&2(n-1-[\frac{u}{2}]) \non\\
	S(n,u>2n-1)&=&-\infty \ ,
\ea\ee
\be\ba{rcl}
\label{U}

	U(n,s<0)&=&2n-3 \non \\
	U(n,s=0)&=&2n-1 \non \\
	U(n,1\leq s\leq 2n-1)&=&2(n-[\frac{s+1}{2}])-1 \non\\
	U(n,s>2n-1)&=&-\infty \ .
\ea\ee
Here, $[q]$ means the largest integer less or equal to $q$.
These upper limits show that the leading $T$ dependence for each order in $g_R$
goes
like

\be
\label{MGlead}
\ba{rcl}
	m &\propto& \lefteqn{m_R\sum_{n=0}^\infty g_R^n T^{2n}} \ , \non\\
	g &\propto& \lefteqn{g_R\sum_{n=0}^\infty g_R^n T^{2n-1}} \ ,
\ea\ee
which could be seen directly from the diagrammatic expansion. We shall now find
out what
values of $k$ that enters in the determination of the $i$-th subleading power
of $T$ in
$M_n$ and $G_n$. Therefore, we look at the terms in \Eqref{MGexp} with $l=2n-i$
and
$u=2n-1-i$, which are given by
\bea
\label{Mi}
	&&\sum_q(\trmr)^qM_{n,q,2n-i}
	=-\sum_{\stackrel{q\leq Q(n,2n-i)}{2n-i\leq L(n,q)}}
	\ \!\!\!\!\!\!\!\!\!\!\!\!' \ \ \
	(\trmr)^q \sum_{k=1}^n\sum_{p=0}^\infty \inv{q}
	M_{n-k,q-1+p-k,2n-i} \non\\
	&&\ \ \ \ \ \ \ \ \ \ \ [\th_{kp}(2+k-p-q-2n+i)+\beta_{kp}(n-k)] \non\\
	+&&\!\!\!\!\sum_{\stackrel{r\leq Q(n,2n-i-r)}{2n-i\leq L(n,r)+r}}
	\ \!\!\!\!\!\!\!\!\!\!\!\!' \ \ \
	\sum_{k=1}^n\sum_{p=0}^\infty \inv{r}
	M_{n-k,r-1+p-k,2n-i-r}
	[\th_{kp}(2+k-p-2n-i)+\beta_{kp}(n-k)] \non\\
\eea
\bea
\label{Gi}
	&&\sum_s(\trmr)^s G_{n,s,2n-1-i}
	=-\sum_{\stackrel{s\leq S(n,2n-1-i)}{2n-1-i\leq U(n,s)}}
	\ \!\!\!\!\!\!\!\!\!\!\!\!' \ \ \
	(\trmr)^s \sum_{k=1}^n\sum_{p=0}^\infty \inv{s}
	G_{n-k,s-1+p-k,2n-1-i} \non\\
	&&\ \ \ \ \ \ \ \ \ \ [\th_{kp}(2+k-p-s-2n+i)+\beta_{kp}(n-k+1)] \non\\
	+&&\!\!\!\!\sum_{\stackrel{r\leq S(n,2n-1-i-r)}{2n-1-i\leq U(n,r)+r}}
	\ \!\!\!\!\!\!\!\!\!\!\!\!' \ \ \
	\sum_{k=1}^n\sum_{p=0}^\infty \inv{r}
	G_{n-k,r-1+p-k,2n-1-i-r}
	[\th_{kp}(2+k-p-2n-i)+\beta_{kp}(n-k+1)] \non\\
\eea
The first term on the RHS is for $q\neq 0$ ($s\neq 0$) and the second is for
$q=0$ ($s=0$).
The general solution to the problem of finding what values of  $k$ contribute
for a given
value of $i$ is rather complicated because of the many special cases that have
to be
considered.  The cases of $i=0,1$ are treated in next section. The way of
solving the
problem for general $i$ is discussed in Appendix B, here I just state the
result.  To resum
the $i$-th subleading powers of $T$ using the TRGE, the $\th$ and $\beta$
functions have
to be calculated to the $k$-th order where $k$ satisfies
\bea
\label{mcond}
\ba{rc}
	m: & k\leq i/2, \ {\rm all\ } p \\
	 & {\rm or\ } p+k\leq   i+1
\ea   \\
\label{gcond}
\ba{rc}
	g: & k\leq i/2, \ {\rm all\ }p \\
	 & {\rm or\ } p+k\leq i+2
\ea
\eea
Remember that the $i$ is defined as $T^{2n-i}$ for $m$ and as $T^{2n-1-i}$ for
$g$.

A discussion of the result is given in Section \ref{concl} and here we just
notice that the
solution of the TRGE, using a perturbatively calculated $\th$ and $\beta$ (\ie
finite $k$),
indeed results in a resummation of leading and subleading factors of $\tmr$.
%
%''''''''''''''''''''''''''''''''''''''''''''''''''''''''''''''''''
\subsection{Special cases when $i=0,1$}
\label{i01}

The most important cases are, of course, the leading and first subleading
contributions.
We start with the leading terms for the mass which go like in \Eqref{MGlead}.
The analysis
in Appendix \ref{appB} gives that only $k=1,\ p=0$ contribute, \ie $\th_{10}$.
That can also
be seen from a diagrammatic expansion of $m^2$; only the 1-loop tadpole give a
$g_RT^2$ dependence, all other diagrams give at most $g_R^nT^{2n-1}$.
Therefore, the
leading $T$ dependence of $m(T)$ is given by
\be
\label{mT}
	m(T)=\sqrt{m_R+g_R\th_{10}(T^2-T_R^2)} \ .
\ee
The next-to-leading order ($i=1$) requires the knowledge of
$\th_{10},\th_{11},\th_{20},\beta_{11}$ and $\beta_{20}$  (from $p+k\leq 2$,
\Eqref{mcond}) which gives a much more complicated TRGE to solve.
\\ \\
Let us continue with the expansion of $g(T)$ which gives a more interesting
result. It turns
out that to leading order ($i=0$) it is necessary to calculate

$\th_{10},\beta_{11}$ and $\beta_{20}$

\footnote{ From \Eqref{gcond} we have $p+k\leq 2$, but $\th_{11}$ and
$\th_{20}$ do not
contribute since the bracket $[2+k-p-s-2n+i]$ vanishes, see \Eqref{Gi}. If we
had included
the $\ln T$ terms the bracket would in general not be of the same form.
However, to
highest order there are no $\ln T$ terms so the conclusion is correct.}.

That is, to get the leading $T$ dependence we must actually do a 2-loop
calculation!
%
% ** lowest order diagrams **
\bfig
\begin{picture}(20000,25000)(-3000,0)
	\thicklines
	\drawline\fermion[\NE\REG](0,0)[2000]
	\drawline\fermion[\NW\REG](\pbackx,\pbacky)[2000]
	\global\advance \pfrontx by 2100
	\put(\pfrontx,\pfronty){\circle{4400}}
	\global\advance \pfronty by 2100
	\put(\pfrontx,\pfronty){\circle*{800}}
	\global\advance \pfronty by -2100
	\global\advance \pfrontx by 2100
	\drawline\fermion[\NE\REG](\pfrontx,\pfronty)[2000]
	\drawline\fermion[\SE\REG](\pfrontx,\pfronty)[2000]
	\global\advance \pfrontx by 6000
	\put(\pfrontx,\pfronty){\makebox(0,0)[l]{
		$\lefteqn{
		=-\frac{3g_R^3}{16\pi^4}\frac{\del^2 F^2_1(T,m_R)}{\del m_R^4}
		F^2_1(T_R,m_R)\approx -\frac{g_R^3 TT_R^2}{256\pi m_R^3}
		}$ }}
	\drawline\fermion[\NE\REG](0,8000)[2000]
	\drawline\fermion[\NW\REG](\pbackx,\pbacky)[2000]
	\global\advance \pfrontx by 2100
	\put(\pfrontx,\pfronty){\circle{4400}}
	\global\advance \pfronty by 3000
	\put(\pfrontx,\pfronty){\circle{2000}}
	\global\advance \pfronty by -3000
	\global\advance \pfrontx by 2100
	\drawline\fermion[\NE\REG](\pfrontx,\pfronty)[2000]
	\drawline\fermion[\SE\REG](\pfrontx,\pfronty)[2000]
	\global\advance \pfrontx by 6000
	\put(\pfrontx,\pfronty){\makebox(0,0)[l]{
		$\lefteqn{
		=\frac{3g_R^3}{16\pi^4}\frac{\del^2 F^2_1(T,m_R)}{\del m_R^4}
		F^2_1(T,m_R)\approx \frac{g_R^3 T^3}{256\pi m_R^3}
		}$ }}

	\drawline\fermion[\NE\REG](0,16000)[2000]
	\drawline\fermion[\NW\REG](\pbackx,\pbacky)[2000]
	\global\advance \pfrontx by 2100
	\put(\pfrontx,\pfronty){\circle{4400}}
	\global\advance \pfrontx by 2100
	\drawline\fermion[\NE\REG](\pfrontx,\pfronty)[2000]
	\drawline\fermion[\SE\REG](\pfrontx,\pfronty)[2000]
	\global\advance \pfrontx by 6000
	\put(\pfrontx,\pfronty){\makebox(0,0)[l]{
		$\lefteqn{
		=\frac{3g_R^2}{4\pi^2}\frac{\del F^2_1(T,m_R)}{\del m_R}
		\approx -\frac{3g_R^2 T}{16\pi m_R}
		}$ }}
	\put(\pfrontx,23000){\makebox(0,0)[l]{
	$\lefteqn{
	=\frac{g_R}{4\pi^2}F^2_1(T,m_R) \approx \frac{g_R T^2}{24}
	}$ }}
	\drawline\fermion[\E\REG](0,23000)[7000]
	\global\advance \pmidy by 1500
	\put(\pmidx,\pmidy){\circle{3000}}
	\global\advance \pfrontx by 12000
\end{picture}
\caption{\it Diagrams contributing to the lowest order $\th$ and $\beta$. The
vertex is
renormalized at zero momentum.}
\label{tb}
\efig
Since the leading order TRGE can be solved explicitly we shall take a closer
look at it to
see what the actual values of  $\th$, $\beta$, $m(T)$ and $g(T)$ are. What we
need to
know is $\th_{10}$, $\beta_{11}$ and $\beta_{20}$, and they are obtained from
the
diagrams in \Figref{tb},
where we use the notation

\be
	F^p_q(T,m)=\int_0^\infty\frac{dk\,k^p}{\omega^q}
	\frac{1}{e^{\omega/T}-1},\ \omega=\sqrt{k^2+m^2} \ .
\ee
{}From these diagrams we find the TRGE
\bea
	\frac{dm}{dT} &=& \frac{gT}{24m} \ ,\\
	\frac{dg}{dT} &=& -\frac{3g^2}{16\pi m}
		+\frac{g^3T^2}{128\pi m^3} \ .
\eea
The last equation can be written as an exact differential and the solution when
$T_R=0$ is

\be
\label{gT}
	g(T)=\frac{g_R}{1+\frac{3g_R}{16\pi}\frac{T}{m(T)}}\ .
\ee
Then the equation for $m(T)$ reads
\be
	\frac{dm}{dT}=\frac{2\pi g_R T}{48\pi m+9g_RT}\ ,
\ee
which, after a change of variables $m(T)=v(T)\, T$, turns into a separable
equations and
can be solved by integration. The implicit solution is
\be
	m^2+\frac{3g_R}{16\pi} mT-\frac{g_R}{24}T^2
	=m_R^2\left(
	\frac{96\pi m+T(\sqrt{81g_R^2+384\pi^2g_R}+9g_R)}
	{96\pi m-T(\sqrt{81g_R^2+384\pi^2 g_R}-9g_R)}\right)
	^{\frac{9g_R}{\sqrt{81g_R^2+384\pi^2 g_R}}}
\ee
In the $T\rightarrow\infty$ limit we find that
\be
	\frac{m(T)}{T}\rightarrow\sqrt{\frac{g_R}{24}
	+(\frac{3g_R}{32\pi})^2} - \frac{3g_R}{32\pi}\ ,
\ee
and
\be
\label{gTlim}
	g(T)\rightarrow g_R\frac{\sqrt{81g_R^2+384\pi^2 g_R}-9g_R}
	{\sqrt{81g_R^2+384\pi^2g_R}+9g_R}\ .
\ee
In order to determine the subleading behaviour of $g(T)$ we would need to
calculate
$\th_{10}$, $\th_{11}$, $\th_{20}$, $\beta_{11}$, $\beta_{12}$, $\beta_{20}$,
$\beta_{21}$
and $\beta_{30}$. Note that the subleading terms can become dominant since the
leading
terms sum up to a constant when $T\rightarrow\infty$ (see Section
\ref{effprop}).
%
% ================================================================
%
\section{Conclusion and outlook}
\label{concl}

The main conclusion of this paper is that the TRGE resums the leading and
subleading
powers of $T$ in much the same way as the $T=0$ RGE resums leading $\ln \mu$
terms. It
is also interesting to notice that a 2-loop calculation is required to get the
leading
behaviour for the coupling constant. This was observed by Funakubo and Sakamoto
in
Ref.\cite{funakubos87} who studied the $O(N)$ model in the limit $N\rightarrow
\infty$
where it is exactly solvable.

It is, of course, not a miracle that some higher order diagrams are determined
in terms of
lower order diagrams. At zero temperature, the leading $\ln \mu$ contribution
comes from
diagrams with many divergent subdiagrams. It is those subdiagrams that are
controlled by
the $\beta$ function. Higher order diagrams with leading $\ln \mu$ terms are
constructed
by multiple insertion of those subdiagrams. At finite temperature there is one
type of loop
that differs from others: the quadratically divergent  tadpole which gives a
factor $T^2$
\cite{dolanj74,fendley87}. All other loops give a factor $T$ (in the high $T$
limit). At the
$n$-th order, if we want the $n$-th subleading contribution to $m(T)$ ($\propto
g_R^n
T^n$) all diagrams to that order contribute.
It is, therefore, natural that we have to calculate the $\th$ and $\beta$
functions to $n$-th
order (actually, even if we want the $(n-1)$-th subleading term we get a $k=n$
contribution
from the condition $p+k\leq i+1$ (\Eqref{mcond}) when $p=0$ and $i=n-1$). We
can briefly
say that the TRGE is a way of resumming the subdiagrams with a $T^2$ dependence
of
which there is only the 1-loop tadpole in the $(\phi^4)_4$-theory.

A difference between the zero and finite $T$  RGE is that the $\beta$ function
at zero $T$
does not depend on $\mu$ in the MS scheme, while the $\beta$ function at finite
$T$
depends on $T$. That makes it harder to prove a leading $T$ resummation and it
works
only because the $T$ dependence is not too strong, only $g_R^nT^n$ and not
$g_R^nT^{2n}$. The reason for that is, as we saw in Section \ref{powtb}, that
the highest
power of $T$ is subtracted off by the counterterms. Then we are back to the
physical
starting point, namely to use the physical $T$ dependent mass already in  the
propagator.
In that way the highest $T$ dependence does not contribute to the $\beta$
function.
\subsection{Effective propagators}
\label{effprop}
The method of using an effective propagator with a mass term $m_R^2+g_RT^2/24$
also
resums the leading powers of $T$ to each order in $g_R$. By definition of the
effective
mass we have the same expression for $m(T)$ when $T_R=0$ (\Eqref{mT}) and from
the
1-loop diagram in \Figref{tb} we get for the coupling constant
\be
\label{gTeffprop}
	g(T)=g_R+\frac{3g_R^2}{32\pi^2}\ln(1+\frac{g_RT^2}{24m_R^2})
	-\frac{3g_R^2}{16\pi}\frac{T}
	{\sqrt{m_R^2+\frac{g_RT^2}{24}}}
\ee
which agrees with \Eqref{gT} to leading order.

Still they predict quantitatively different behaviours at very high
temperature. The
logarithmic term in \Eqref{gTeffprop}, that comes from the renormalization at
$T=0$, is
formally of lower order in $T/m_R$ and should be neglected. Then
\Eqref{gTeffprop} gives
a negative effective 4-point coupling at sufficiently high $T$ if
$g_R>32\pi^2/27$,
 which indicates an instability even though it can be stabilized by higher
n-point functions
(such as the 6-point function). On the other, hand the logarithmic term is
actually dominant
at high $T$ and gives an increasing $g(T)$.
 The TRGE result in \Eqref{gTlim} does not show any of these behaviours. The
difference
between the effective propagator approach and the TRGE is in the subleading
terms, and
it is clear that if the leading terms sum up to something finite when
$T\rightarrow\infty$ the
subleading terms can become dominant. That is also the reason why $m(T)$ and
$g(T)$
are found to diverge at a critical temperature in Ref.\cite{funakubos87} but
not here.

A problem with the effective propagator is that it leads to $T$ dependent
infinities in the
tadpole diagram in \Figref{tadpole} when the mass is explicitly $T$ dependent.
It occurs in
subleading terms and is subtracted away if all diagrams to that order are
included
\cite{kislingerm76}. Also, in \cite{fendley87} this was pointed out as a defect
of not treating
subleading terms consistently while in \cite{parwani91} it was considered to be
legitimate
with such $T$ dependent infinities. In my point of view the bare parameters of
the theory
has nothing to do with the thermal state of the system and must, therefore, be
temperature
independent.
The TRGE is, however, {\it derived} from the $T$ independence of the bare
parameters
and satisfy this criterion by definition. It is a subject for further study to
understand
significance of subleading terms in the TRGE approach.
\subsection{Other theories}
\label{othther}
As we now understand, the method of resummation using the TRGE works because
the
leading temperature corrections can be accurately approximated by a $T$
dependent
mass term. In the $(\phi^4)_4$-theory it is particularly simple since the
tadpole does not
depend on the external momentum. The possibility of extending the result of
this paper to
other theories is related to the possibility of including the leading $T$
dependence in the
propagator, \ie to expand about the physical excitations at finite temperature.
For instance,
the $(\phi^3)_6$-theory has a momentum dependent 1-loop correction to the mass.
It turns
out, however, that the leading $T^2$ part of that correction is momentum
independent and
the TRGE should work even for that theory.
\\ \\
In QCD, the gluon polarization tensor has a $T^2$ term which {\it is} momentum
dependent
and cannot be approximated by a constant \cite{kalashnikovk80,weldon82}. This
indicates
that the TRGE, as described here, does not yield a consistent resummation. It
may be
possible to generalize the TRGE to include a momentum dependent mass term. The
choice of RC determines the propagator and we have only considered  finite
changes in a
momentum independent mass by renormalizing at a fixed momentum. In principle,
one can
allow for a momentum dependent mass term. The reason for {\it not} doing so is
that it
makes calculations more complicated, but if the physical excitations cannot be
well
approximated by free particles with $T$ dependent mass, one has to use
something better.
A perturbation method using effective propagators and vertices to include the
$T^2$
contributions was developed in \cite{braatenp90}. The analysis in this paper
indicates that
a 2-loop calculation would be necessary also in QCD to get a TRGE which resums
the
leading powers of $T$ for the coupling constant.
\\ \\
I want to thank Drs. Umezawa and Sakamoto for discussions and The Theoretical
Physics
Institute at University of Alberta for their hospitality during my stay. I also
thank The
Swedish Institute for financial support.
%
%++++++++++++++++++++++++++++++++++++++++++++++++++++++++++++++++++
%
\appendix
\section{Appendix}
\label{appA}

We out-line the determination of the functions $Q$,$L$,$S$, and $U$ defined in
Section
\ref{powmg}. From \Eqref{qMsG} we see that
\be
\label{Qmax}
	Q(n,l) = \max_{k,p} [Q(n-k,l)+1-p+k]
\ee
where $1\leq k\leq n$ and $p\geq 0$. If the result is less or equal to zero we
must also
check \Eqref{q0cond} which gives a lower limit
\be
	Q(n,l)\geq 0 \ {\rm if} \ l\leq L(n,r)+r \ ;\ r\neq 0 \ .
\ee
Since the determination of $Q(n,l)$ only involves $Q(n',l)$ with $n'<n$ we can
solve the
problem iteratively in $n$ with initial condition
\be
	Q(0,0)=0\ ,\ Q(0,l\neq 0)=-\infty \ .
\ee
Each time we determine $k,p$ in \Eqref{Qmax} we must be sure that the factor in
brackets
in \Eqref{qMsG} does not vanish for these values. It does not pose any problem
here but it
does when calculating $S$ and $U$.

Similarly $L$ is then determined by
\be\ba{rcl}
	L(n,q\neq 0)&=&\max_{k,p}[L(n-k,q-1+p-k)]  \non\\
	L(n,0)&=&\max_{r\neq 0} [L(n,r)+r] \ ,
\ea\ee
with the same initial condition as for $Q$.
\\ \\
The procedure is the same for $S$ and $U$ but we find that
\be
	S(1,0)=\max_p [S(0,0)+1-p+1] = 2\ ,
\ee
occurs for $k=1,\ p=0$. This term is, however, absent in  \Eqref{qMsG} since
$\beta_{10}=0$ and $1+k-p-s-u=0$. Therefore, $S(1,0)=1$ for $k=1,\ p=1$. That
is the
major difference between $Q,L$ and $S,U$. As we can see from \Eqref{qMsG}
$S(n,0)$
and $U(n,0)$ has to be treated separately from $S(n,1\leq u\leq 2n-1)$ and
$U(n,1\leq
s\leq 2n-1)$.
%
% +++++++++++++++++++++++++++++++++++++++++++++++++++++++++++++++
%
\section{Appendix}
\label{appB}

In order to determine which values of $k$ and $p$ that occurs in the summations
in
\Eqsref{Mi}{Gi} we  look at the first and second term separately.

In the first term in \Eqref{Mi} the summation over $q$ runs through all values
satisfying
\be\ba{rcl}
	q&\leq& Q(n,2n-i)  \non\\
	 2n-i&\leq& L(n,q)\ .
\ea\ee
For each such $q$ the terms that contribute to the $k$ and $p$ summations
satisfy
\be\ba{rcl}
	q-1+p-k&\leq& Q(n-k,2n-i)  \non\\
	2n-i &\leq& L(n-k,q-1+p-k)\ .
\ea\ee
Finally, the factors in brackets in \Eqsref{Mi}{Gi} must be non-zero. This
analysis gives that
for $i>2n$ there is no restriction on $k$ and $p$ (other than $1\leq k\leq n$
and $p\geq 0$,
of course) and for $2\leq i\leq 2n$ we have $k\leq i/2$. There is no
contribution from the
first term when $i=0,1$.
\\ \\
The summation range for the second term is determined by
\be\ba{rcl}
	r&\leq& Q(n,2n-i-r)  \non\\
	 2n-i&\leq& L(n,r)+r \ ,
\ea\ee
and the terms in the  $k,p$ sums satisfy
\be\ba{rcl}
	r-1+p-k&\leq& Q(n-k,2n-i-r)  \non\\
	2n-i-r &\leq& L(n-k,r-1+p-k) \ .
\ea\ee
We find that this term contributes whenever $p+k\leq i+1$.
\\ \\
In a similar way we check the summation in \Eqref{Gi} where in the first term
the
summation runs over all $s$ satisfying
\be
\ba{rcl}
	s&\leq& S(n,2n-1-i) \\
	2n-1-i&\leq& U(n,s) \ ,
\ea
\ee
and the $k,p$ summation ranges over
\be
\label{Gkp}
\ba{rcl}
	s-1+p-k&\leq& S(n-k,2n-1-i) \\
	2n-1-i &\leq& U(n-k,s-1+p-k) \ .
\ea
\ee
We find, using \Eqsref{S}{U}, almost the same conditions as for $M$, \ie $k\leq
i/2$ and
$p+k\leq i+1$. However, the expression for $S(0,u)$ and $U(0,s)$ do not follow
the general
form and give exceptions for $n=k$ in the RHS of \Eqref{Gkp}. This gives an
extra term
when $k=n$ and $p\leq i+2-n$. The highest $n$ (=$k$) for which this term can
contribute is
when $p=0$, \ie $n=k=i+2$.
\\ \\
In summary, we have found that for each iteration of the solution of
\Eqsref{Mi}{Gi} (if we
imagine an iterative solution in $n$) the terms contributing are the ones
satisfying
\be
\label{kpcond}
\ba{rl}
	& k\leq i/2,\ {\rm all\ } p \ \
	{\rm (also\ } 1\leq k\leq n {\rm )} \\
	{\rm or } & p+k\leq i+1 \ .
\ea
\ee
In addition, in the $(i+2)$-th order in \Eqref{Gi} there is an extra term with
\be
\label{kpcond2}
	p+k\leq i+2\ .
\ee
We conclude that, in order to get the $i$-th subleading term of the
$T_R/m_R$-expansion
of $m(T)$ we need to compute the $\th$ and $\beta$ functions to the $(i+1)$-th
order
(respectively the $(i+2)$-th order for $g(T)$). Actually, we only need the
leading
$T$-dependent part of the highest order diagram when calculating $\th$ and
$\beta$. The
\Eqsref{kpcond}{kpcond2} tell more precisely what is needed. An explicit
example for
$i=0$ is given in Section \ref{i01}.
%
% -------------------- bibliography ---------------------

%\newpage
%\bibliographystyle{unsrt}
%\addcontentsline{toc}{chapter}{References}
%\bibliography{}

%
%
\end{document}